\documentclass[letterpaper,preprintnumbers,prd,twocolumn,nofootinbib,nobibnotes,showpacs]{revtex4}
\usepackage{amsfonts}
\usepackage{mathrsfs}
\usepackage{epsfig}
\usepackage{graphicx}%
\usepackage{dcolumn}
\usepackage{amsmath}
\usepackage{color}
\usepackage{color}
\makeatletter
\def\btt#1{\texttt{\@backslashchar#1}}%
\DeclareRobustCommand\bblash{\btt{\@backslashchar}}%
\makeatother
\begin{document}

\title{An exact black hole spacetime with scalar field and its shadow together with quasinormal modes }
\author{Shuang Yu}\email{yushuang@nao.cas.cn} \affiliation{ Key Laboratory of Computational Astrophysics, National Astronomical Observatories, Chinese Academy of Sciences, Beijing 100101, China}
\affiliation{School of Astronomy and Space Sciences, University of Chinese Academy of Sciences,
No. 19A, Yuquan Road, Beijing 100049, China}\author{Changjun Gao}\email{gaocj@bao.ac.cn}\affiliation{ Key Laboratory of Computational Astrophysics, National Astronomical Observatories, Chinese
Academy of Sciences, Beijing 100101, China}
\affiliation{School of Astronomy and Space Sciences, University of Chinese Academy of Sciences,
No. 19A, Yuquan Road, Beijing 100049, China}

\date{\today}

%%%%%%%%%%%%%%%%%%%%%%%%%%%%%%%%%%%%%%%%%%%%%%%%%%%%%%%%%%%%%%%%%%%%%%%%%%
\begin{abstract}
We find an exact black hole solution with a minimally coupled scalar field. The corresponding spacetime has two horizons and one of the them is the black hole event horizon and the other is the cosmic horizon. In this sense, the solution is analogous to the Schwarzschild-de Sitter (or anti-de Sitter) spacetime.  We investigate the thermodynamics and construct the first law of thermodynamics. At the same time, we make a study on the shadow and quasinormal modes of this black hole solution.
\end{abstract}


%\keywords{}

\maketitle

%%%%%%%%%%%%%%%%%%%%%%%%%%%%%%%%%%%%%%%%%%%%%%%%%%%%%%%%%%%%%%%%%%%%%%%%%%
\section{Introduction}
\label{sec:1}
The first solution of the Einstein equations with self-interacting minimally coupled scalar field was investigated by Fisher \cite{Fisher}. This solution brings us many interesting properties, for example, the logarithmical divergence problem of total energy for the static field. The subsequent studies on the Einstein-scalar theories lead to the construction of no-hair theorem on black holes or particle-like asymptotically flat spacetimes \cite{Rosen:1966,Chase:1970,Bekenstein:1972,Teitelboim:1972,Adler:1978,Heusler:1992,Sudarsky:1995,Bekenstein:1995,Galtsov:2001,Bronnikov:2001,Bronnikov:2002}. However, there are other black holes \cite{Torii:2001,Nucamendi:2003,Martnez:2004,Anabalon:2012,Radu:2005} which can evade no-hair theorem. These solutions remain important because the standard energy conditions assumed in the proof of no-hair theorem can often be violated in modern physics. Even when the interactions between black holes and matter fields \cite{Stefanov:2008,Doneva:2010,Cardoso:2013a,Cardoso:2013b} are considered, the no-hair theorem can be violated. Furthermore, there is another key assumption in the proof of no-hair theorem. Namely, the spacetime of a black hole is assumed to be asymptotically flat and the scalar field asymptotically vanishing. Given up this assumption, we will frequently obtain other new black hole solutions. Actually, Sotiriou and Zhou \cite{sot:14} have pointed out that, in principle, one scalar field could always have a nontrivial configuration although without the black hole carrying an extra (independent) scalar charge. This is sometime
referred to as ``hair of the second kind''. In this paper, we report we find an exact black hole solution with the minimally coupled scalar field. The spacetime is not asymptotically flat and thus it evades the constraints from no-hair theorem. We shall investigate the solution from the following theoretical and  observational aspects.

In the first place, basing on the theoretical considerations, we shall study its thermodynamics. We know, ever since the discovery of Hawking radiation \cite{Hawking1975}, black holes become thermodynamic systems from the gravitational ones which have absolute zero temperature. Black holes thus possess internal energy \cite{Smarr1973,Bardeen1973}, entropy \cite{Bekenstein1973} and other thermal quantities. It is then found that black hole entropy is proportional to the area of event horizon and the black hole temperature is proportional to the surface gravity on event horizon \cite{Hawking1975,Bardeen1973}. Afterwards, the four laws of black hole mechanics turn out to be four laws of black hole thermodynamics \cite{Bekenstein1973,Hawking1972,Hawking1975,Bardeen1973}. Then black holes are not gravitational systems with zero temperature, but thermodynamical systems with finite temperatures. But up to now, the interpretation of these four laws in the viewpoint of micro-physics remains an open question. In this regard, an important connection between gravity and thermodynamics is found by Jacobson \cite{Jacoboson1995}. Nowadays, black hole thermodynamics has grown into an important research field in modern physics \cite{Hayward1994,Hayward1998,Jacobson1995,Eling2004,Hayward2004,Ashtekar2002,Gong2007,Rogatko2007,Padmanabhan2005,Padmanabhan2002,Cai2007,Chakraborty2015,Padmanabhan2010,
Gibbons1997,Paranjape2006,Wei2009,Durka2019,Hareesh2019,Mahdieh2019}. In view of these considerations, we shall investigate the thermodynamics of our black hole solution.

Secondly, as the first observational aspect, we shall compute the shadow of the black holes. Recent observations of black holes from event horizon telescope provide
the first image of black hole in the center of galaxy $\textrm{M}87$ \cite{event1,event2}. Though the first image
 is not enough to identify the black hole geometry, it confirms straightforwardly that the black holes
 does exist in our universe. Therefore, it is of great importance to improve the resolution of the telescope in the observations
 \cite{Goddi2016} and calculate the forms of shadows cast by black holes and black-hole mimickers \cite{Cunha2016,wang2019,Younsi2016,Cunha2017,Bisnovatyi2017,Bisnovatyi2018,Perlick2018,
 Tsupko2018,amir2017,xu2018,wang2017a,Tsukamoto2018,Abdujabbarov2017,wang2017b,wang2018,hennigar2018,
 Dokuchaev2017,Shaikh2018,Mizuno2018,Dokuchaev2018,Amir2019,ovgun2018,Huang2018,Wei2019a,Abdikamalov2019,
 Held2019,Shaikh2019,Kumar2019,wei2019b,Konoplya2019}. In view of these points, we shall calculate the form of shadow for our black hole solution.

Thirdly, as the second observational aspect, we shall calculate its quasinormal modes. As is known that the detections of gravitational waves at LIGO and VIRGO \cite{ligo,virgo} mark the beginning of the astronomical era of gravitational waves. The dominating contribution to gravitational waves when a black hole is perturbed comes from the quasinormal modes. It is a long period of damping proper oscillations whose frequencies only depend on the parameters of the black hole, such as mass, charge and angular momentum. Thus it is called the ``character sound'' of black holes. Making perturbations to a black hole can be performed in two ways. One is by adding matter fields of different spins to the black hole spacetimes and the other is by perturbing the black hole metric itself. Either way, the black hole undergo damped oscillations at the intermediate stage with complex frequencies. The real part of the frequency describes the oscillation rate and the imaginary part describes the damping rate. These oscillations are called quasi-normal modes. In this paper, we shall adopt the first method to calculate the quasi-normal modes.

The paper is organized as follows. In Sec. II, we seek for the exact black hole solution with minimally coupled scalar field. In Sec. III, we analyze the energy conditions for the scalar field. In Sec. IV, we make an analysis on the thermodynamics of the black hole solution. In Sec. V we calculate the shadow of black hole. Sec. VI is devoted to computation of quasinormal modes.
Finally, the conclusion and discussion is given in Sec. VII. Throughout this paper, we adopt the system of units in which $G=c=\hbar=1$ and the metric signature
$(-,\ +,\ +,\ +)$.

\section{black hole solution}
We consider a scalar field with the Lagrangian
\begin{equation}
\mathscr{L}=\frac{1}{2}\partial_{\mu}\phi\partial^{\mu}\phi+V\left(\phi\right)\;,
\end{equation}
where
\begin{equation}
V\left(\phi\right)=\frac{1}{l^2}\cdot\left(\frac{1}{2}-ke^{-\frac{2}{3}\sqrt{6}\phi}\right)e^{\sqrt{6}\phi}\;,
\end{equation}
is the self-interaction potential. This is a double-exponential potential which is widely used in cosmology \cite{Jarv:04}. The constant $l$ denotes some length and $k$ is dimensionless. The corresponding energy momentum tensor is
\begin{equation}
T_{\mu\nu}=-\partial_{\mu}\phi\partial_{\nu}\phi+g_{\mu\nu}\left(\frac{1}{2}\partial_{\alpha}\phi\partial^{\alpha}\phi+V\right)\;.
\end{equation}
Then the Einstein equations are
\begin{eqnarray}
G_{\mu\nu}=-\partial_{\mu}\phi\partial_{\nu}\phi+g_{\mu\nu}\left(\frac{1}{2}\partial_{\alpha}\phi\partial^{\alpha}\phi+V\right)\;.
\end{eqnarray}
The equation of motion for $\phi$ is
\begin{eqnarray}
\nabla_{\mu}\nabla^{\mu}\phi-\frac{\partial V}{\partial\phi}=0\;.
\end{eqnarray}
In order to obtain the static and spherically symmetric black hole solution, we find it is convenient to assume the metric in the form with a conformal factor $N(x)$:
\begin{eqnarray}
ds^2=N\left(x\right)\left[-U\left(x\right)dt^2+\frac{1}{U\left(x\right)}dx^2+x^2d\Omega^2\right]\;,
\end{eqnarray}
where $d\Omega^2$ is the line element for two dimensional unit sphere and $x$ is the radial coordinate. By solving equations Eq.~(4) and Eq.~(5), we find a solution which is given by

\begin{eqnarray}
N&=&\frac{xl^2}{\left(4Q-x\right)^3}\;,\\
U&=&-\left(1-\frac{2Q}{x}\right)^2+\frac{1}{192Q^2}\left(3-2k\right)x^2\;,\\
\phi&=&\frac{\sqrt{6}}{2}\ln\left(4Q-x\right)-\frac{\sqrt{6}}{2}\ln x\;,
\end{eqnarray}
with $Q$ an integration constant. We note that $Q$ is not a physical parameter for the black hole. In fact, it can be eliminated
by coordinate transformations. To this end, we make coordinate transformation form $x$ to $r$ by putting
\begin{eqnarray}
x=\frac{4Qr^2}{l^2+r^2}\;.
\end{eqnarray}
Then the metric becomes
\begin{eqnarray}
&&ds^2=-\frac{l^2}{192Q^2}\left(-\frac{3l^2}{r^2}+\frac{6r^2}{l^2}-\frac{2kr^6}{l^6}\right)dt^2\nonumber\\&&
+\frac{48r^6}{l^6}\left(-\frac{3l^2}{r^2}+\frac{6r^2}{l^2}-\frac{2kr^6}{l^6}\right)^{-1}dr^2+\frac{r^6}{l^4}d\Omega^2\;.
\end{eqnarray}
The factor in front of the line element $d\Omega^2$ is not $r^2$ but $r^6/l^6$. With the help of this trivial difference, the expression of the metric looks considerable simple. $Q$ can be absorbed by $t$. Then the metric is

\begin{eqnarray}\label{eq:metric1}
&&ds^2=-\left(-\frac{3l^2}{r^2}+\frac{6r^2}{l^2}-\frac{2kr^6}{l^6}\right)dt^2\nonumber\\&&
+\frac{48r^6}{l^6}\left(-\frac{3l^2}{r^2}+\frac{6r^2}{l^2}-\frac{2kr^6}{l^6}\right)^{-1}dr^2+\frac{r^6}{l^4}d\Omega^2\;.
\end{eqnarray}
Now we see $Q$ is in the absence. We note that there are only two physical parameters, $l$ and $k$ in the metric while without the mass parameter $M$. Therefore this is not general but particular solution and it does not have the Schwarzschild limit. Furthermore, it does not has the Minkowski limit. However, we find the causal structure can be Schwarzschild-de Sitter or Schwarzschild-anti-de Sitter. The reason for this point is that, we find $l$ plays the role of mass and $k$ the role of cosmological constant in section IV. The spacetime structure analysis will be shown in the next.

For this spacetime, there exists a {space-like} singularity at $r=0$ and the scenarios of horizons are as follows:
 (1). When $0<k<\frac{3}{2}$ and $l>0$, there are two horizons at
\begin{eqnarray}\label{eq:hori}
r_{+}&=&l\left[\frac{1}{2k}\left(3+\sqrt{9-6k}\right)\right]^{\frac{1}{4}}\;,\\
r_{-}&=&l\left[\frac{1}{2k}\left(3-\sqrt{9-6k}\right)\right]^{\frac{1}{4}}\;.
\end{eqnarray}
Here $r_{+}$ and $r_{-}$ denote the cosmic and black hole horizon, respectively.
The causal structure of the spacetime is exactly the same as the Schwarzschild-de Sitter solution.

(2). When $k<0$ and $l>0$, the cosmic horizon disappears and there is only one black hole solution. In this case, the corresponding causal structure is exactly the same as the Schwarzschild-anti-de Sitter solution.

(3). When $k=0$ and $l>0$, we are left with unique black hole horizon
 \begin{eqnarray}
r_{EH}=2^{-\frac{1}{4}}l\;.
\end{eqnarray}
Namely, the size of black hole is proportional to the length scale $l$. This  spacetime is asymptotically neither Minkowski nor de Sitter (or anti-de Sitter).

(4). When $k=\frac{3}{2}$ and $l>0$, the black hole horizon and cosmic horizon coincide and the causal structure is exactly the same as extreme Schwarzschild-de Sitter solution.

(5). When $k>\frac{3}{2}$ and $l>0$, there would be no horizons and the singularity is naked. The corresponding causal structure is  the same as Schwarzschild-de Sitter solution but with a naked singularity.
\section{energy conditions}
In this section, we analyze the energy conditions when the black hole has two horizons. To this end, we put $l=1$ and $k=1$. Then the two horizons
locate at $r_-=0.89$ and $r_+=1.24$, respectively.
The energy conditions include:

(1) Null Energy Condition(NEC), $\rho+p_{r} \geq 0 \text { and } \rho+p_{T} \geq 0$\;;

(2) Weal Energy Condition(WEC), $\rho \geq 0, \rho+p_{r} \geq 0, \quad \rho+p_{T} \geq 0$\;;

(3) Strong Energy Condition(SEC), $\rho+p_{r}+2 p_{T} \geq 0, \rho+p_{r} \geq 0, \text { and } \rho+p_{T} \geq 0$\;;

(4) Dominant Energy Condition(DEC), $\rho \geq 0, \rho \pm p_{r} \geq 0, \text { and } \rho \pm p_{T} \geq 0$\;.

\begin{figure}[h]
\begin{center}
\includegraphics[width=9cm]{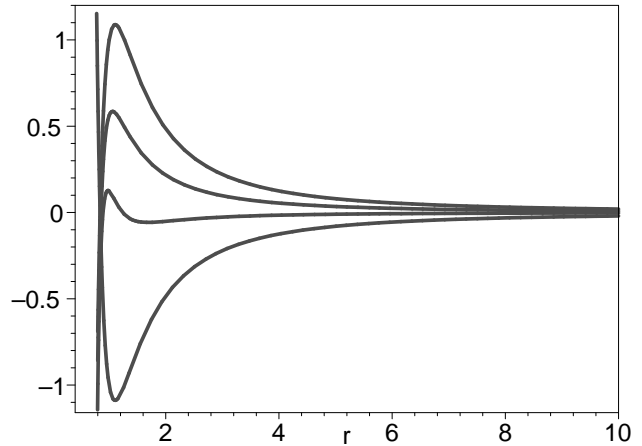}
\caption{The variation of $\rho-p_r$, $\rho$, $\rho+p_r$ and $\rho+p_r+2p_T$ with respect to $r$ from up to down. $\rho$ and $\rho-p_r$ are always positive outside the black hole horizon. $\rho+p_r\geq 0$ holds exactly between the black hole horizon and the cosmic horizon. $\rho+p_r+2p_T<0$ holds outside the black hole event horizon.
}\label{fig:ec}
\end{center}
\end{figure}

Given the metric, one can calculate the energy density $\rho$, the radial pressure $p_r$ and tangent pressure $p_T$ as follows
\begin{eqnarray}
&&\rho =-\frac{2 r^{4} l^{4}-14 k r^{8}+3 l^{8}}{128\pi r^{10}}\;,\\&&
p_{r}=-\frac{3 l^{8}-14 r^{4} l^{4}+18 k r^{8}}{128\pi r^{10}}\;,\\&&
p_{T}=-\rho\;.
\end{eqnarray}
In Fig.~\ref{fig:ec} we plot the variation of $\rho$, $\rho-p_r$, $\rho+p_r$ and $\rho+p_r+2p_T$ with respect to $r$ from up to down.
It shows that $\rho$ and $\rho-p_r$ are always positive outside the black hole horizon. As for $\rho+p_r$, we have $\rho+p_r\geq 0$ exactly between the black hole horizon and the cosmic horizon. Lastly, we have $\rho+p_r+2p_T<0$ outside the black hole event horizon.
Thus we conclude that the NEC, WEC and DEC are always satisfied between the black hole horizon and the cosmic horizon. The SEC is always violated.

\section{thermodynamics}
In this section, we construct the first law of thermodynamics for this spacetime. Since there are in general two horizons, we should seek for the corresponding laws with respect to each horizon. To this end, let's start from the derivation of thermodynamical quantities of the black hole. In the first place, the entropy of the two horizons are
given by
\begin{eqnarray}\label{eq:entro}
S_{\pm}&=&\frac{\pi r_{\pm}^6}{l^4}\;,
\end{eqnarray}
because of Bekenstein-Hawking law, $S=A/4$. Here the plus and minus in ``$\pm$'' denote the cosmic and black hole horizon, respectively. To calculate
the temperatures of the two horizons, we resort to the method of ``imaginary-time period" \cite{pad:2010}. To this end, we make Taylor series expansion of the metric in the vicinity of horizons
\begin{eqnarray}\label{eq:metric}
&&ds^2\simeq g_{00_{,r}}|_{r=r_{\pm}}\cdot\left(r-r_{\pm}\right)dt^2\nonumber\\&&
+\frac{48r_{\pm}^6}{l^6}\left[-g_{00_{,r}}|_{r=r_{\pm}}\cdot\left(r-r_{\pm}\right)\right]^{-1}dr^2+\frac{r_{\pm}^6}{l^4}d\Omega^2\;.
\end{eqnarray}
We see it reduces to the Rindler metric in the vicinity the horizons in the $t-r$ plane with the surface
gravity
\begin{eqnarray}
\kappa_{\pm}=-\frac{l^3}{8\sqrt{3}r_{\pm}^3}{g_{00_{,r}}|_{r=r_{\pm}}}\;.
\end{eqnarray}
Then, an analytic continuation to imaginary time allows one to
identify the temperatures associated with the horizons to be
\begin{eqnarray}\label{eq:temp1}
T_{\pm}=\frac{|\kappa_{\pm}|}{2\pi}=\frac{\sqrt{3}l\left(l^2+r_{\pm}^2\right)\left(l+r_{\pm}\right)|\left(l-r_{\pm}\right)|}{2\pi r_{\pm}^6}\;.
\end{eqnarray}
As for the energy $\mathscr{E}_{\pm}$, the pressure $P_{\pm}$ and the thermodynamical volume $V_{\pm}$, we find they can be written as
\begin{eqnarray}\label{eq:temp2}
\mathscr{E}_{\pm}&=&\sqrt{3}\left(\l-\frac{r_{\pm}^4}{l^3}\right)\;,\\
P_{\pm}&=&\frac{\sqrt{3}}{6\pi}\left(\frac{9l}{r_{\pm}}-\frac{r_{\pm}^3}{l^3}\right)\;,\\
V_{\pm}&=&2\pi r_{\pm}\;.
\end{eqnarray}
After expressing the above thermodynamic quantities in terms of $r_{\pm}$ and $k$, we arrive at the first law of thermodynamics
\begin{eqnarray}\label{eq:temp3}
\delta\mathscr{E}_{\pm}&=T_{\pm}\delta S_{\pm}+ V_{\pm}\delta P_{\pm}\;.
\end{eqnarray}
Observing the expressions for horizons, Eq.~(\ref{eq:hori}) and that for energy and pressure in Eq.~(23) and Eq.~(24),
we find that the black hole energy is proportional to $l$ and the pressure is uniquely determined by $k$. Remembering the role of mass and cosmological
constant in the thermodynamics of Schwarzschild-anti-de Sitter solution \cite{teit:1985,brown:1988}, we recognize $l$ plays the role of mass and $k$ the role of cosmological constant.

\section{black hole shadow}
In this section, we calculate the black hole shadow as seen by a static observer at $r=r_{o}$. It is apparent we should have $r_{-}\leq r_{o}\leq r_{+}$.
Because the spacetime is static and spherically symmetric, without the loss of generality, we can locate the observer in the equatorial plane $\theta_{o}=\pi/2$.
So it is sufficient for us to study null geodesics in the equatorial
plane. The geodesics of light in the equatorial plane are determined by the
Lagrangian
\begin{eqnarray}
\mathscr{L}=\frac{1}{2}\left[g_{00}\dot{t}^2+g_{11}\dot{r}^2+g_{22}\dot{\varphi}^2\right]\;,
\end{eqnarray}
where $g_{00}, g_{11}, g_{22}$ are the metric components of line element Eq.~(\ref{eq:metric1}). Then the Euler-Lagrange equation with respect to $t$ and $\varphi$ give us two constants of motion:
\begin{eqnarray}\label{eq:two}
E=-g_{00}\dot{t}\;,\ \ \ \ \  L=g_{22}\dot{\varphi}\;,\ \ \ \ \
\end{eqnarray}
with $E$ and $L$ denoting the energy and angular momentum of a photon, respectively.
For null geodesics we have

\begin{eqnarray}\label{eq:null}
\mathscr{L}=0\;.
\end{eqnarray}
Taken into account Eqs.~(\ref{eq:two}), the null geodesics equation Eq.~(\ref{eq:null}) can be written as
\begin{eqnarray}\label{eq:eom}
\left(\frac{dr}{d\varphi}\right)^2=\frac{\left(l^2E^2+2kL^2\right)r^8-6l^4L^2r^4+3l^8L^2}{48l^4r^2L^2}\;.
\end{eqnarray}
By demanding the equations $dr/d\varphi=0$ and $d^2r/d\varphi^2$ we would acquire the circular null geodesic at
\begin{eqnarray}
r_c=l\;,
\end{eqnarray}
and that the dimensionless constant $k$ for this
circular geodesic should satisfy
\begin{eqnarray}\label{eq:k}
k=\frac{3}{2}-\frac{l^2E^2}{2L^2}\;.
\end{eqnarray}
It is found this circular geodesic is unstable. The reason is that a
slight perturbation of the initial direction for the light ray would lead it to escape from the
circle at $r=r_c$ and enters either the black hole horizon $r_{-}$ or the cosmic horizon $r_{+}$.
Taking three spatial dimensions into account, we would have an unstable null sphere at $r=r_c$.

In order to construct the shadow we consider the light
rays that are emitted by the observer at
($r=r_o$). The light rays leave the
observer at an angle $\theta$ with respect to the radial line
\begin{eqnarray}
\tan\theta=\frac{r^3ld\varphi}{4\sqrt{3}r^3\left(-\frac{3l^2}{r^2}+\frac{6r^2}{l^2}-\frac{2kr^6}{l^6}\right)^{-\frac{1}{2}}dr}\mid_{r=r_{o}}\;.
\end{eqnarray}
By substituting Eq.~(\ref{eq:eom}) into it, we get
\begin{eqnarray}
\tan^2\theta=\frac{L^2\left(-3l^8+6l^4r_{o}^4-2kr_{o}^8\right)}{\left(l^2E^2+2kL^2\right)r_{o}^8-6l^4L^2r_{o}^4+3l^8L^2}\;,
\end{eqnarray}
or
\begin{eqnarray}
&&\sin^2\theta=\frac{L^2\left(-3l^8+6l^4r_{o}^4-2kr_{o}^8\right)}{r_{o}^8l^2E^2}\;.
\end{eqnarray}

\begin{figure}[h]
\begin{center}
\includegraphics[width=9cm]{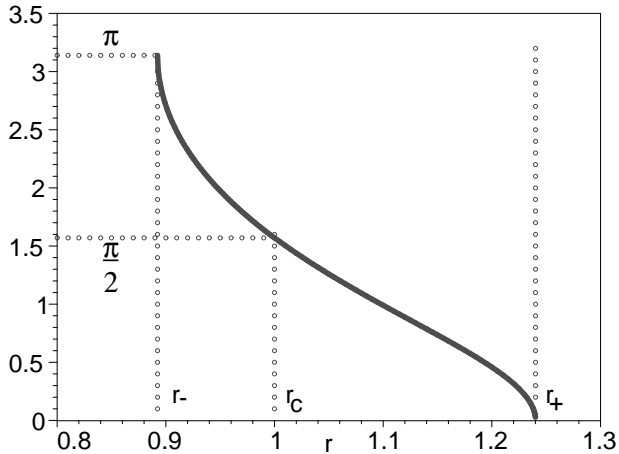}
\caption{The variation of angular radius $\theta_s$ with respect to the position of observer. It shows when $r_o=r_{+}$, the total sky is bright and when $r_o=r_{-}$, the total sky is dark. But when $r_o=r_{c}$, half of the sky is bright. For simplicity, we have put $E=1, l=1$ and $L=1$.
}\label{fig:theta}
\end{center}
\end{figure}

The boundary of the shadow is determined by light rays that spiral asymptotically
towards circular null geodesics at $r=r_c$. Therefore, we substitute $k$ with Eq.~(\ref{eq:k}) and arrive at
\begin{eqnarray}
\sin^2\theta_{s}=1-\frac{3l^6L^2}{r_{o}^8E^2}+\frac{6l^2L^2}{r_{o}^4E^2}-\frac{3L^2}{l^2E^2}\;.
\end{eqnarray}
Observing the equation, we find $\theta_s=\pi/2$ when $r=r_{c}$. This means half of the sky is dark if the observer locates at $r_{o}=r_c$. On the other hand,
if the observer locates at $r_{o}=r_{\pm}$, we have $\theta_s=0, \pi$, respectively. This means the total sky is bright or dark, respectively.
In Fig.~(\ref{fig:theta}) we plot the variation of angular radius $\theta_{s}$ in terms of the position of observer.

\section{quasinormal modes}

In this section we study the quasinormal modes generated by
the propagation of a test minimally coupled massless scalar field. To this end, we start from the well-known Klein-Gordon equation
\begin{eqnarray}
\nabla^2\Psi=0\;,
\end{eqnarray}
which is the general perturbation equation for the massless scalar field in the curved spacetime.
Here $\nabla^2$ is the four dimensional Laplace operator and $\Psi$ the massless scalar field. For arbitrary static spherically symmetric black hole spacetimes, we have derived the effective potential in Ref.(\cite{We2019}). Thus we can write the radial perturbation equation
\begin{eqnarray}
\frac{d^2\Phi}{d r_{\ast}^2}+\left(\omega^2-V\right)=0\;,
\end{eqnarray}
where the tortoise coordinate $r_{\ast}$ is defined by
\begin{eqnarray}
 {{{r}_{*}}}\equiv\int \frac{4\sqrt{3}r^5l^3}{-3 l^8+6 r^4 l^4-2kr^8}{dr}\;,
\end{eqnarray}
and the effective potential is given by
\begin{equation}
\begin{aligned}
\text{V} = & \frac{1}{16\left(l^{12} r^{14}\right)} \left(3 l^{8}-6 r^{4} l^{4}+2 k r^{8}\right) \left({-16 r^{12} l^{6} m^{2}}\right. \\ & \left. {-16 r^{6} l^{10} s^{2}  -16 r^{6} l^{10} s-24 r^{6} l^{10}+16 r^{10} l^{10}}\right. \\ & \left.{+27 l^{16}-108 l^{12} r^{4}+36 l^{8} k r^{8}+108 r^{8} l^{8}}\right. \\ & \left.{-72 r^{12} l^{4} k+12 k^{2} r^{16} }\right)\;.
\end{aligned}
\end{equation}
Here $s=0,1,2,3,\cdot\cdot\cdot$ is the multipolar number.

\begin{figure}[h]
\begin{center}
\includegraphics[width=9cm]{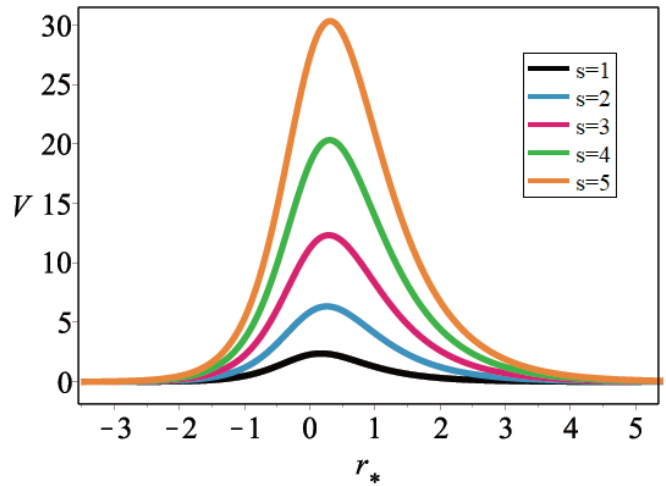}
\caption{The effective potential $V(r_{\ast})$ as a function of the tortoise coordinate $r_\ast$ when $l=1, k=1$ and $s=5,4,3,2,1$, from up to down.}\label{pot}
\end{center}
\end{figure}

\begin{figure}[h]
\begin{center}
\includegraphics[width=9cm]{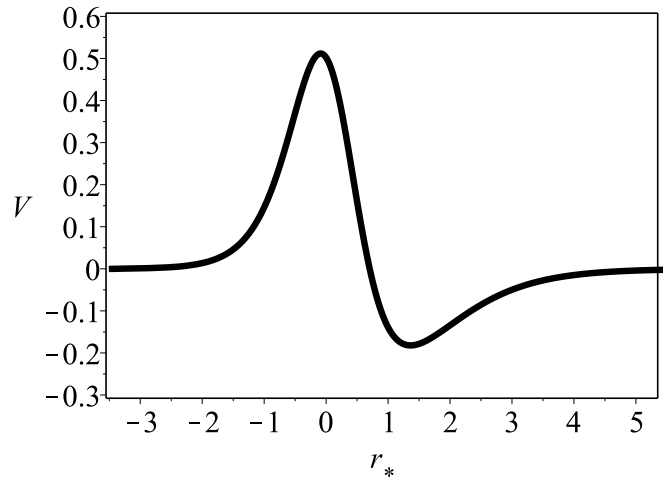}
\caption{The effective potential $V(r_{\ast})$ as a function of the tortoise coordinate $r_\ast$ when $l=1, k=1$ and $s=0$. In this case, a negative gap emerges which would lead to an eikonal instability.}\label{pot0}
\end{center}
\end{figure}
In Fig.~\ref{pot}, we plot the effective potential $V$ as the function of the tortoise coordinate $r_\ast$ when $l=1, k=1$ and $s=1,2,3,4,5$.
The event horizon and spatial infinity locates at $r_{\ast}=-\infty$ and $r_{\ast}=\infty$, respectively. From the figure we see, except for a maximum for the potential, it always asymptotically vanishes on both the event horizon and the spatial infinity. With the increasing of multipolar number $s$, the height of the potential is increased. In Fig.~\ref{pot0}, we plot the effective potential $V(r_{\ast})$ with respect to $r_\ast$ when $l=1, k=1$ and $s=0$.
In this case, a negative gap emerges in the potential and, according to \cite{taka:2009}, this would lead to an eikonal instability and the WKB method  \cite{will:1985,will:1987,iyer:1987} for the calculation of quasinormal modes is not yet applicable.
The third-order WKB approximation, a numerical and perhaps the most popular method, is devised by Schutz, Will and Iyer \cite{will:1985,will:1987,iyer:1987}.
This method has been used extensively in evaluating quasinormal frequencies of various black holes. For an incomplete list see \cite {quasi,Kokkotas:99,Berti:09} and references therein.

The quasinormal frequencies are given by
\begin{eqnarray}
\omega^2=V_0+\Lambda\sqrt{-2V_0^{''}}-i\nu\left(1+\Omega\right)\sqrt{-2V_0^{''}}\;,
\end{eqnarray}
where $\Lambda$ and $\Omega$ are

\begin{eqnarray}
\Lambda&=&\frac{1}{\sqrt{-2V_0^{''}}}\left\{\frac{V_0^{(4)}}{V_0^{''}}\left(\frac{1}{32}+\frac{1}{8}\nu^2\right)
\right.\nonumber\\&&\left.-\left(\frac{V_0^{'''}}{V_0^{''}}\right)^2\left(\frac{7}{288}+\frac{5}{24}\nu^2\right)\right\}\;,\\
\Omega&=&\frac{1}{\sqrt{-2V_0^{''}}}\left\{\frac{5}{6912}\left(\frac{V_0^{'''}}{V_0^{''}}\right)^4\left(77+188\nu^2\right)
\right.\nonumber\\&&\left.-\frac{1}{384}\left(\frac{V_0^{'''2}V_0^{(4)}}{V_0^{''3}}\right)\left(51+100\nu^2\right)
\right.\nonumber\\&&\left.+\frac{1}{2304}\left(\frac{V_0^{(4)}}{V_0^{''}}\right)^2\left(67+68\nu^2\right)\right.\nonumber\\&&\left.
+\frac{1}{288}\left(\frac{V_0^{'''}V_0^{(5)}}{V_0^{''2}}\right)\left(19+28\nu^2\right)\right.\nonumber\\&&\left.-\frac{1}{288}\left(\frac{V_0^{(6)}}{V_0^{''}}
\left(5+4\nu^2\right)\right)\right\}\;,
\end{eqnarray}
and

\begin{eqnarray}
\nu=n+\frac{1}{2}\;,\ \ \ \ V_0^{(s)}=\frac{d^sV}{dr_{\ast}^s}|_{r_{\ast}=r_p}\;,
\end{eqnarray}
$n$ is overtone number and $r_p$ corresponds to the peak of the effective
potential. It is pointed that \cite{car:04} that the accuracy of the WKB method depends on the multiple
number $s$ and the overtone number $n$. The WKB approach is consistent with the numerical method very well provided that  $s>n$. Therefore we shall
present the quasinormal frequencies of scalar perturbation for $n=0$ and $s=1,2,3,4,5$, respectively.

In table I, we give the fundamental quasinormal frequencies for $l=6$ and different $k$ and $s$.  From the table we see that with the increasing of the coupling constant $k$, both the real part and the imaginary part of the frequencies are decreasing, which means that with the increasing of $k$, both the oscillating and the decay of scalar perturbation become slower and slower. On the other hand, with the increasing of multipolar number $s$, the real part increases and imaginary part decreases. This reveals that the oscillating of scalar perturbation becomes faster and faster while the decay of perturbation becomes slower and slower. We do not give the quasinormal modes for $k<0.6$. The reason for this is that the potential always has a negative gap which is similar to Fig.~\ref{pot0}. The presence of the gap makes the WKB method inapplicable. In table II, we give the fundamental quasinormal frequencies for $k=1$ but varying $l$ and $s$. From the table we see, the effect of $l$ on the frequencies is similar to $k$, while the effect of $s$ remains the same.

\section{Conclusion and Discussion}\label{sec:8}

 In conclusion, we find and investigate an exact black hole solution with a minimally coupled scalar field. The potential of the scalar field consists of two exponential potentials which are widely used in cosmology. Two parameters $l$ and $k$ appear in the potential. They are also present in the metric. Since there is no mass parameter in the metric, we conclude it is not a general, but particular solution of the Einstein equations. The search for the general solution with mass parameter is an open question.

 On the other hand, the solution is not asymptotically flat in space. This guarantees it does not violate the no-hair theorem. In general, there are two horizons in this spacetime. One is for the black hole event horizon and the other for the cosmic horizon. This is similar to the Schwarzschild-de Sitter solution. It is found $l$ plays the role of black hole mass and $k$ the role of cosmological constant. The investigation on thermodynamics tells us the total energy of black hole is proportional to $l$. By calculating the entropy, energy, pressure, temperature and thermodynamical volume, we construct the first law of thermodynamics. We find the pressure is uniquely determined by the coupling constant $k$ which is much like the role of cosmological constant in the Schwarzschild-anti-de Sitter thermodynamics.

 Whereafter, we study the black hole from two observational aspects, namely, the aspects of black hole shadow and quasinormal modes. We prove that the unstable circular geodesic locates at $r_{c}=l$ while it is independent on $k$.  This gives another clear physical meaning to $l$. It is shown that if an observer stands on the sphere of $r_{c}=l$, he will see half of the sky is dark. On the other hand, if the observer stands on the cosmic horizon or black hole event horizon, he will see the sky is totally bright or dark, respectively.

 In the aspect of quasinormal modes, we compute the modes with varying $k,s$ and $l,s$, respectively. The results show $l$ and $k$ have the same effect on the modes. Concretely, with the increasing of the coupling constant $k$ (or $l$), both the real part and the imaginary part of the frequencies are decreasing. This means with the increasing of $k$ (or $l$), both the oscillation and the decay of scalar perturbation become slower and slower. For the effect of multipolar number $s$, it is found with the increasing of $s$, the real part increases and imaginary part decreases. This reveals that the oscillation of scalar perturbation becomes faster and faster while the decay of perturbation becomes slower and slower.

\section*{Acknowledgments}
We are grateful to the referee for the expert review. This work is partially supported by China Program of International ST Cooperation 2016YFE0100300
, the Strategic Priority Research Program ``Multi-wavelength Gravitational Wave Universe'' of the
CAS, Grant No. XDB23040100, the Joint Research Fund in Astronomy (U1631118), and the NSFC
under grants 11473044, 11633004 and the Project of CAS, QYZDJ-SSW-SLH017.

\begin{table*}[h]
\begin{center}
\begin{tabular}[b]{ccccccc}
 \hline \hline
 $k$&$\omega (s=1)$\;&\;$\omega (s=2)$\;&\;$\omega (s=3)$\;&\;$\omega (s=4)$\;&\;$\omega(s=5)$\;& \;$\omega (s=6)$\; \\ \hline
0.6&0.322968-0.156014I&0.545780-0.128101I&0.773023-0.119460I&0.998883-0.116146I&1.223900-0.114567I&1.448461-0.113698I \\
0.8&0.281799-0.128031I&0.482949-0.108084I&0.683346-0.102944I&0.882221-0.101048I&1.080432-0.100152I&1.278309-0.099661I \\
1.0&0.236766-0.101309I&0.408559-0.088839I&0.577708-0.085886I&0.745679-0.084787I&0.913152-0.084262I&1.080369-0.083972I \\
1.2&0.181247-0.074714I&0.315162-0.067701I&0.446442-0.066049I&0.576746-0.065419I&0.706611-0.065114I&0.836239-0.064943I \\
1.4&0.101528-0.041205I&0.180133-0.038496I&0.256443-0.037864I&0.331964-0.037612I&0.407127-0.037499I&0.482096-0.037431I \\
\hline \hline
\end{tabular}
\end{center}
\caption{The fundamental ($n=0$) quasinormal frequencies of scalar field for the black hole when $l=6$.}
\end{table*}

\begin{table*}[h]
\begin{center}
\begin{tabular}[b]{ccccccc}
 \hline \hline
 $l$&$\omega (s=1)$\;&\;$\omega (s=2)$\;&\;$\omega (s=3)$\;&\;$\omega (s=4)$\;&\;$\omega(s=5)$\;& \;$\omega (s=6)$\; \\ \hline
2&0.699950-0.278887I&1.223054-0.254182I&1.732042-0.250270I&2.236446-0.249106I&2.739081-0.248688I&3.240847-0.248543I \\
4&0.354329-0.150036I&0.612630-0.132297I&0.866477-0.128252I&1.118472-0.126770I&1.369699-0.126073I&1.620533-0.125694I \\
6&0.236766-0.101309I&0.408559-0.088839I&0.577708-0.085886I&0.745679-0.084787I&0.913152-0.084262I&1.080369-0.083972I \\
8&0.177719-0.076318I&0.306456-0.066797I&0.433296-0.064515I&0.559267-0.063662I&0.684870-0.063253I&0.810280-0.063025I \\
10&0.142228-0.061178I&0.245178-0.053500I&0.346642-0.051649I&0.447417-0.050956I&0.547897-0.050623I&0.648226-0.050437I\\
\hline \hline
\end{tabular}
\end{center}
\caption{The fundamental ($n=0$) quasinormal frequencies of scalar field for the black hole when $k=1$.}
\end{table*}

\newcommand\ARNPS[3]{~Ann. Rev. Nucl. Part. Sci.{\bf ~#1}, #2~ (#3)}
\newcommand\AL[3]{~Astron. Lett.{\bf ~#1}, #2~ (#3)}
\newcommand\AP[3]{~Astropart. Phys.{\bf ~#1}, #2~ (#3)}
\newcommand\AJ[3]{~Astron. J.{\bf ~#1}, #2~(#3)}
\newcommand\GC[3]{~Grav. Cosmol.{\bf ~#1}, #2~(#3)}
\newcommand\APJ[3]{~Astrophys. J.{\bf ~#1}, #2~ (#3)}
\newcommand\APJL[3]{~Astrophys. J. Lett. {\bf ~#1}, L#2~(#3)}
\newcommand\APJS[3]{~Astrophys. J. Suppl. Ser.{\bf ~#1}, #2~(#3)}
\newcommand\JHEP[3]{~JHEP.{\bf ~#1}, #2~(#3)}
\newcommand\JMP[3]{~J. Math. Phys. {\bf ~#1}, #2~(#3)}
\newcommand\JCAP[3]{~JCAP {\bf ~#1}, #2~ (#3)}
\newcommand\LRR[3]{~Living Rev. Relativity. {\bf ~#1}, #2~ (#3)}
\newcommand\MNRAS[3]{~Mon. Not. R. Astron. Soc.{\bf ~#1}, #2~(#3)}
\newcommand\MNRASL[3]{~Mon. Not. R. Astron. Soc.{\bf ~#1}, L#2~(#3)}
\newcommand\NPB[3]{~Nucl. Phys. B{\bf ~#1}, #2~(#3)}
\newcommand\CMP[3]{~Comm. Math. Phys.{\bf ~#1}, #2~(#3)}
\newcommand\CQG[3]{~Class. Quantum Grav.{\bf ~#1}, #2~(#3)}
\newcommand\PLB[3]{~Phys. Lett. B{\bf ~#1}, #2~(#3)}
\newcommand\PRL[3]{~Phys. Rev. Lett.{\bf ~#1}, #2~(#3)}
\newcommand\PR[3]{~Phys. Rep.{\bf ~#1}, #2~(#3)}
\newcommand\PRd[3]{~Phys. Rev.{\bf ~#1}, #2~(#3)}
\newcommand\PRD[3]{~Phys. Rev. D{\bf ~#1}, #2~(#3)}
\newcommand\RMP[3]{~Rev. Mod. Phys.{\bf ~#1}, #2~(#3)}
\newcommand\SJNP[3]{~Sov. J. Nucl. Phys.{\bf ~#1}, #2~(#3)}
\newcommand\ZPC[3]{~Z. Phys. C{\bf ~#1}, #2~(#3)}
\newcommand\IJGMP[3]{~Int. J. Geom. Meth. Mod. Phys.{\bf ~#1}, #2~(#3)}
\newcommand\IJMPD[3]{~Int. J. Mod. Phys. D{\bf ~#1}, #2~(#3)}
\newcommand\IJMPA[3]{~Int. J. Mod. Phys. A{\bf ~#1}, #2~(#3)}
\newcommand\GRG[3]{~Gen. Rel. Grav.{\bf ~#1}, #2~(#3)}
\newcommand\EPJC[3]{~Eur. Phys. J. C{\bf ~#1}, #2~(#3)}
\newcommand\PRSLA[3]{~Proc. Roy. Soc. Lond. A {\bf ~#1}, #2~(#3)}
\newcommand\AHEP[3]{~Adv. High Energy Phys.{\bf ~#1}, #2~(#3)}
\newcommand\Pramana[3]{~Pramana.{\bf ~#1}, #2~(#3)}
\newcommand\PTP[3]{~Prog. Theor. Phys{\bf ~#1}, #2~(#3)}
\newcommand\APPS[3]{~Acta Phys. Polon. Supp.{\bf ~#1}, #2~(#3)}
\newcommand\ANP[3]{~Annals Phys.{\bf ~#1}, #2~(#3)}
\newcommand\RPP[3]{~Rept. Prog. Phys. {\bf ~#1}, #2~(#3)}

\end{document}